\documentclass[preprint,11pt]{artcle2}
\usepackage[top=2cm,bottom=2cm,left=2.5cm,right=2cm]{geometry}
\usepackage[utf8]{inputenc}
\usepackage[T1]{fontenc}
\usepackage{bm}
\usepackage{caption}
\usepackage{subcaption}
\usepackage{graphicx}
\usepackage{amsmath,bm}
\usepackage{mathtools}
\usepackage{stmaryrd}
\usepackage{tabularx}
\usepackage{graphicx}
\usepackage{xcolor}
\usepackage[left, pagewise]{lineno}
\usepackage{relsize}
\biboptions{numbers,sort&compress}
\usepackage{hyperref}
\hypersetup{
    colorlinks=true,
    linkcolor=blue,      
    urlcolor=cyan,
    }
\makeatletter
\DeclareRobustCommand\bigop[1]{%
  \mathop{\vphantom{\sum}\mathpalette\bigop@{#1}}\slimits@
}
\newcommand{\bigop@}[2]{%
  \vcenter{%
    \sbox\z@{$#1\sum$}%
    \hbox{\resizebox{\ifx#1\displaystyle.9\fi\dimexpr\ht\z@+\dp\z@}{!}{$\m@th#2$}}%
  }%
}
\makeatother
\newcommand{\Assembly}{\DOTSB\bigop{\mathrm{A}}}
\usepackage{amssymb}

\newcommand{\boldface}[1]{\boldsymbol{#1}}  

\newcommand{\bfb}{\boldface{b}}

\newcommand{\bfd}{\boldface{d}}
\newcommand{\bfe}{\boldface{e}}
\newcommand{\bff}{\boldface{f}}

\newcommand{\bfh}{\boldface{h}}

\newcommand{\bfj}{\boldface{j}}

\newcommand{\bfm}{\boldface{m}}
\newcommand{\bfn}{\boldface{n}}

\newcommand{\bfq}{\boldface{q}}
\newcommand{\bfr}{\boldface{r}}

\newcommand{\bft}{\boldface{t}}
\newcommand{\bfu}{\boldface{u}}
\newcommand{\bfv}{\boldface{v}}
\newcommand{\bfw}{\boldface{w}}

\newcommand{\bfB}{\boldface{B}}
\newcommand{\bfC}{\boldface{C}}
\newcommand{\bfD}{\boldface{D}}
\newcommand{\bfE}{\boldface{E}}
\newcommand{\bfF}{\boldface{F}}

\newcommand{\bfH}{\boldface{H}}

\newcommand{\bfK}{\boldface{K}}

\newcommand{\bfR}{\boldface{R}}

\newcommand{\bfU}{\boldface{U}}

\newcommand{\bfepsilon}{\boldsymbol{\varepsilon}}

\newcommand{\bfkappa}{\boldsymbol{\kappa}}

\newcommand{\bfmu}{\boldsymbol{\mu}}

\newcommand{\bfsigma}{\boldsymbol{\sigma}}

\newcommand{\bfphi}{\boldsymbol{\phi}}

\newcommand{\calH}{\mathcal{H}}
\newcommand{\calI}{\mathcal{I}}

\newcommand{\calL}{\mathcal{L}}

\newcommand{\calP}{\mathcal{P}}

\newcommand{\calU}{\mathcal{U}}
\newcommand{\calV}{\mathcal{V}}
\newcommand{\calW}{\mathcal{W}}

\newcommand{\dsC}{\mathbb{C}}

\newcommand{\dsR}{\mathbb{R}}

\newlength{\boxwidth}
\def\btheorem{\begin{theorem}}
\def\etheorem{\end{theorem}}
\def\blemma{\begin{lemma}}
\def\elemma{\end{lemma}}
\def\bproposition{\begin{proposition}}
\def\eproposition{\end{proposition}}
\def\bcorollary{\begin{corollary}}
\def\ecorollary{\end{corollary}}
\def\bdefinition{\begin{definition}}
\def\edefinition{\end{definition}}
\def\bexample{\begin{example}}
\def\eexample{\end{example}}
\def\bremark{\begin{remark}}
\def\eremark{\end{remark}}

\def\fg{\boldsymbol}

\def\vdot{\fg{\cdot}}

\newcommand{\be}{\begin{equation}}
\newcommand{\ee}{\end{equation}}
\newcommand{\beq}{\begin{eqnarray}}
\newcommand{\eeq}{\end{eqnarray}}
\newcommand{\bem}{\begin{multline}}
\newcommand{\eem}{\end{multline}}
\newcommand{\ba}{\begin{align}}
\newcommand{\ea}{\end{align}}

\begin{document}

\begin{frontmatter}

\title{A discontinuous Galerkin method based isogeometric analysis framework for flexoelectricity in micro-architected dielectric solids}

\author{Saurav Sharma}
\author{Cosmin Anitescu}
\author{Timon Rabczuk\corref{cor1}}
\ead{timon.rabczuk@uni-weimar.de}
\cortext[cor1]{Corresponding author}

\address{Institute of Structural Mechanics, Bauhaus University Weimar, Weimar 99423, Germany}
\begin{abstract}
Flexoelectricity -- the generation of electric field in response to a strain gradient -- is a universal electromechanical coupling, dominant only at small scales due to its requirement of high strain gradients. This phenomenon is governed by a set of coupled fourth-order partial differential equations (PDEs), which require $C^1$ continuity of the basis in finite element methods for the numerical solution. While Isogeometric analysis (IGA) has been proven to meet this continuity requirement due to its higher-order B-spline basis functions, it is limited to simple geometries that can be discretized with a single IGA patch. For the domains, e.g., architected materials, requiring more than one patch for discretization IGA faces the challenge of $C^0$ continuity across the patch boundaries. Here we present a discontinuous Galerkin method-based isogeometric analysis framework, capable of solving fourth-order PDEs of flexoelectricity in the domain of truss-based architected materials. An interior penalty-based stabilization is implemented to ensure the stability of the solution. The present formulation is advantageous over the analogous finite element methods since it only requires the computation of interior boundary contributions on the boundaries of patches. As each strut can be modeled with only two trapezoid patches, the number of $C^0$ continuous boundaries is largely reduced. Further, we consider four unique unit cells to construct the truss lattices and analyze their flexoelectric response. The truss lattices show a higher magnitude of flexoelectricity compared to the solid beam, as well as retain this superior electromechanical response with the increasing size of the structure. These results indicate the potential of architected materials to scale up the flexoelectricity to larger scales, towards achieving universal electromechanical response in meso/macro scale dielectric materials.

\end{abstract}


\end{frontmatter}

\section{Introduction}
\label{sec:Intro}
Electromechanical coupling, i.e., inter-conversion of mechanical and electrical energies, has proven to be vital in many engineering applications -- ranging from structural health monitoring \cite{na2018SHM,park2007SHM,raghavan2005SHM}, control \cite{sharma2020Control,shivashankar2020control}, and energy harvesting \cite{panda2022Energyharvesting,sharma2022Energyharvesting,sezer2021Energyharvesting}, to biomedical implants \cite{liu2023advances,chen20233d} and devices \cite{panwar2017design,roy2019self}. The most widely studied mechanism that enables this coupling is piezoelectricity, which stems from the separation of charges-- and thus generation of polarization-- in non-centrosymmetric materials upon deformation and conversely the deformation of materials upon application of an electric field. However, the main bottlenecks of piezoelectricity are its limited presence across material classes, brittleness, and chemical toxicity of prominent lead-based piezoelectric materials. On the contrary, flexoelectricity -- the generation of electric polarization in response to strain gradient -- is a phenomenon ubiquitously present in all dielectric materials, providing a superior electromechanical response at micro/nano scales. Flexoelectricity originates from the breaking of mirror symmetry at the molecular level, facilitated by non-uniform strain. As a consequence, this phenomenon requires high magnitudes of strain gradients to be able to break the mirror symmetry at the crystal level and thus is dominated by piezoelectricity at macro and meso scales. However, at the smaller scales, the roles are reversed, since a non-uniform deformation exhibits sharp strain gradients due to the smaller size of the structure. This mechanism of generating electrical signal from non-uniform deformation has been found to be the driving mechanism of many biological processes, e.g., regeneration of bones \cite{vasquez2018flexoelectricity,witt2023modelling}, hearing \cite{deng2019collusion,breneman2009hair}, plant-based sponges \cite{jiang2023giant}, and biological membranes \cite{ahmadpoor2013apparent,deng2014flexoelectricity,petrov2002livingmembranes}. The source of the electrical activity in these biological systems is mainly their intricate micro-architecture, which ensures a strain gradient inside the structure and thus triggers flexoelectricity-enabled electromechanical response. While the advancements in additive manufacturing technologies make it possible to realize such intricate structures at small scales \cite{dou2018ultralow,mao20193d}, owing to the higher order coupling between mechanical and electrical fields, the computational analysis of these structures is rather challenging.

Micro-architected materials also referred to as metamaterials, allow tailoring their behavior and achieving exceptional and exotic properties, such as negative Poisson's ratio \cite{prall1997properties, felsch2023controlling}, tailored anisotropy \cite{zheng2021data, van2023inverse, nguyen2020three, ahn2018topology}, and extreme strengths \cite{wang2020quasiperiodic, oh2016quasi}, through designed micro-structures of their building blocks, known as unit cells. Among various types of lattices based on one-dimensional and two-dimensional structural elements, truss-based lattices are the most common choice for unit cells of mechanical metamaterials, mainly due to their ease of analysis, design, and fabrication. While the potential of truss lattice metamaterials for tailoring mechanical \cite{Zheng2023}, and thermal properties \cite{Li2023-pn} has been extensively investigated in the past, the electromechanical property space is relatively less explored in this context. Although the systematic arrangement of individual structural elements for tailoring piezoelectric properties at meso scales has been investigated \cite{shi2019architected, sugino2020analytical}, analysis of electromechanical coupling in architected materials at micro scales is complex, due to size effects and gradient phenomena arising in mechanical and electrical physics. Thus treatment of electromechanical coupling in micro-architected solids requires taking flexoelectricity into account. Evaluation of flexoelectricity in complex domains of micro lattices -- composed of discrete individual domains --  requires additional considerations in computational modeling.

Flexoelectricity is mathematically defined by a system of fourth-order coupled partial differential equations (fourth-order elliptic problem) and thus demands at least \textit{C}\textsuperscript{1} continuity of the basis used to approximate the solution. Classical finite element (FE) formulations, based on \textit{C}\textsuperscript{0} Lagrange polynomial basis functions, lack the continuity required for flexoelectricity due to discontinuous first derivatives across the element boundaries \cite{hughes2012finite}. \textit{C}\textsuperscript{1} continuous FE formulations are available for simple one-dimensional (1D) problems, such as those based on Hermite shape functions for Euler-Bernoulli beam problem, but are complex for 2D and 3D problems and become almost infeasible for coupled multiphysics problems with complex domains. While mixed formulations have been employed for solving such problems with \textit{C}\textsuperscript{0} continuous basis \cite{mao2016mixed, deng2017mixed, tian2021collocation}, by taking primary variables and their derivatives as the degrees of freedom and eliminating the requirement of higher order continuity, the increased number of variables makes them computationally expensive. For example, for bending of thin structural elements, e.g., beams, plates, and shells, rotational degrees of freedom are considered as separate variables in the FE formulations of Mindlin-Reissner plate theory or first-order shear deformation theory \cite{reddy1999theory,naghdi1973theory}, to avoid the stronger continuity requirements. While these approaches introduce additional variables, the alternate rotation (gradient) free approaches require rigorous mathematical efforts for \textit{C}\textsuperscript{1} continuous basis. In this context, Isogeometric analysis (IGA) introduced by Hughes et al. \cite{hughes2005isogeometric} has proven to be a promising approach in dealing with both the issues of higher order continuity and additional variables, by employing non-uniform rational B-splines (NURBS) as its basis functions. IGA has been successfully deployed to solve higher order problems -- including strain gradient elasticity \cite{makvandi2018isogeometric}, rotation-free thin structural elements \cite{cottrell2006isogeometric}, and flexoelectricity \cite{ghasemi2017level,sharma2020geometry}. Most of the literature so far has been focused on solving these problems in simple domains (e.g., rectangular, square, and cubic domains), which can be modeled as a single patch using NURBS objects. However, the micro-architectures, the subject of the present study, require multi-patch discretization to model their geometry. Although, NURBS basis functions are \textit{C}\textsuperscript{$\infty$} over a patch, they run into the same problem of discontinuous derivatives across patch boundaries as the traditional Lagrangian basis functions of FE formulations. This is addressed in the recent literature by developing special multi-patch surfaces to implement $C^1$ continuity in a strong sense in shell structures \cite{farahat2023(a)isogeometric,farahat2023(b)isogeometric}. Engel et al. \cite{engel2002continuous} introduced a Discontinuous Galerkin (DG) method-based approach to circumvent the discontinuity of derivatives on the interior boundaries of elements. This approach has been recently applied (termed interior penalty method therein) for flexoelectricity in finite element \cite{ventura2021c0, balcells2022c0}, and immersed boundary hierarchical B-spline-based \cite{codony2019immersed, barcelo2024computational} formulations. 

Here, we use a DG method-based IGA approach for flexoelectricity to enforce the higher order continuity across element boundaries by introducing weighted residuals of jumps of derivatives across the patch boundaries, in the weak form. To ensure the stability of the solutions, a stabilization term is added to the weak form, involving a mesh-dependent stabilization (penalty) parameter. Although the present DG method-based IGA approach is similar to DG-based FE methods in the sense that it also requires additional consideration at the patch boundaries, it is superior to its FE counterpart since only a few boundaries are \textit{C}\textsuperscript{0} continuous. This leads to less computational time in implementation, due to strong form continuity of derivatives in most of the domain and weakly-enforced continuity at only patch boundaries, which are considerably less than the element boundaries in FEA, since here each strut of the architected materials is discretized with only two trapezoid patches.

The outline of the rest of the article is as follows: In section \ref{sec:Theory}, the theoretical framework of flexoelectricity is presented, revisiting the phenomenological theory and the constitutive model. Next, the variational formulation and incorporation of the discontinuous Galerkin method within the framework of flexoelectric theory and $C^0$ continuous approximation spaces are discussed. Using this variational formulation, section \ref{sec:IGA} presents the implementation of discontinuous Galerkin method based on the interior penalty approach. Section \ref{sec:Validation} presents the convergence studies and validation of the present formulation against analytical results from the literature. In section \ref{sec:Results} simulation studies are conducted to evaluate the direct and converse flexoelectric effects in truss lattice-based dielectric materials -- in compression and bending modes. Lastly, section \ref{sec:Conclusion} concludes the study.

\section{Theoretical framework}
\label{sec:Theory}
\subsection{Constitutive model of flexoelectricity}
This section provides a brief summary of the theoretical model of flexoelectric effect in dielectric solids in the presence of piezoelectricity. Initial phenomenological frameworks for flexoelectricity were introduced by Kogan \cite{Kogan1964}, and Harris \cite{harris1965mechanism} in centrosymmetric crystals as the generation of electric polarization proportional to strain gradient. Later, the (converse) flexoelectric effect in dielectric solids was incorporated in the electrical enthalpy density functional by Mindlin \cite{mindlin1968polarization} as a contribution of polarization gradient to the enthalpy density -- extending Toupin's linear theory of electromechanical interaction in dielectrics. More recently, comprehensive theoretical models of the flexoelectric coupling have been provided in references \cite{sharma2010piezoelectric, codony2021mathematical}, including both direct and converse effects in the electrical enthalpy density as
\be
\label{eq:1}
\calH(\boldsymbol{\epsilon},\boldsymbol{\nabla\epsilon},\boldsymbol{E},\boldsymbol{\nabla E})=\frac{1}{2}\dsC_{ijkl}\epsilon_{ij}\epsilon_{kl}+\frac{1}{2}h_{ijklmn}\epsilon_{ij,k}\epsilon_{lm,n}-\frac{1}{2}\kappa_{ij}E_iE_j-e_{ikl}E_i\epsilon_{kl}+f_{ijkl}E_i\epsilon_{jk,l}+d_{ijkl}E_{i,j}\epsilon_{kl},
\ee
where $\bm{\dsC}$, $\bfh$, and $\bfkappa$ are the fourth-order elasticity tensor, the sixth-order strain gradient elasticity tensor, and the second-order dielectric tensor, respectively. The electromechanical coupling tensors $\bfe$, $\bff$, and $\bfd$ denote third-order piezoelectric tensor, fourth-order direct flexoelectric tensor, and fourth-order converse flexoelectric tensor, respectively. The strain gradient elasticity tensor is written in terms of $\bm{\dsC}$ and a length scale parameter $\calL$. See \ref{sec:appendixA} for details of the material tensors. The state variables chosen in writing the above expression are strain tensor $\boldsymbol{\epsilon}$ and $\bfE$, defined in terms of primary variables displacement $\bfu$, and electric potential $\bm{\phi}$ as

\be
\begin{aligned}
\epsilon_{ij}&=\epsilon_{ji}=\frac{1}{2}(u_{i,j}+u_{j,i}), \text{and}\\
E_i&=-\phi_{,i}.
\end{aligned}
\ee
where a \textit{comma} (,) followed by a subscript denotes a spatial derivative. As shown in \cite{sharma2010piezoelectric}, equation \eqref{eq:1} can be expressed in terms of only one flexoelectric coefficient as
\be
\label{eq:3}
\calH(\boldsymbol{\epsilon},\boldsymbol{\nabla\epsilon},\boldsymbol{E})=\frac{1}{2}\dsC_{ijkl}\epsilon_{ij}\epsilon_{kl}+\frac{1}{2}h_{ijklmn}\epsilon_{ij,k}\epsilon_{lm,n}-\frac{1}{2}\kappa_{ij}E_iE_j-e_{ikl}E_i\epsilon_{kl}-\mu_{ijkl}E_i\epsilon_{jk,l}  ,
\ee
where flexoelectric coefficients' tensor $\bfmu=\bfd-\bff$. By definition, Cauchy stress $\boldsymbol{\hat{\sigma}}$, and the double stress $\Tilde{\bfsigma}$ can be written as
\be
\label{eq:4}
\begin{aligned}
\hat{\sigma}&_{ij}=\frac{\partial\calH}{\partial\epsilon_{ij}}=\dsC_{ijkl}\epsilon_{kl}-e_{ijk}E_k, \text{and} \\
\Tilde{\sigma}&_{ijk}=\frac{\partial\calH}{\partial\epsilon_{ij,k}}=h_{ijklmn}\epsilon_{lm,n} - \mu_{ijkl}E_{l}.
\end{aligned}
\ee
Similarly, the electric displacement $\boldsymbol{\hat{\bfD}}$ and higher order electric displacement $\boldsymbol{\Tilde{\bfD}}$ can be written as,
\be
\label{eq:5}
\begin{aligned}
  \hat{D}&_i=-\frac{\partial\calH}{\partial E_i}= \kappa_{ij}E_j+ e_{ikl}\epsilon_{kl}+\mu_{ijkl}\epsilon_{jk,l}, \text{and} \\
  \Tilde{D}&_{ij}=-\frac{\partial\calH}{\partial E_{i,j}}=0.
\end{aligned}
\ee
Using equations \eqref{eq:4} ans \eqref{eq:5}, the physical stress and the physical electric displacement can be written as

\be
\label{eq:6}
\begin{aligned}
   \sigma_{ij}&=\hat{\sigma}_{ij}-\Tilde{\sigma}_{ijk,k}=\dsC_{ijkl}\epsilon_{kl}-h_{ijklmn}\epsilon_{kl,mn} -e_{ijk}E_k + \mu_{ijkl}E_{k,l}, \text{and} \\
    D_i&=\hat{D}_i-\Tilde{D}_{ij,j}=\kappa_{ij}E_j+e_{ikl}\epsilon_{kl} +\mu_{ijkl}\epsilon_{jk,l}
\end{aligned}
\ee
Equations \eqref{eq:5} and \eqref{eq:6} collectively define the material constitutive law of a dielectric material with piezoelectric and flexoelectric couplings. 

\subsection{Variational formulation and the boundary value problem}
The total free enthalpy of the systems with physical domain $\Omega \subset \dsR^d$ can be written as,
\be
\label{eq:7}
\Pi [\bfu,\bfphi] = \int\limits_\Omega \calH(\bfu,\bfphi) \mathrm{d\Omega} - \calW^{ext},
\ee
and the corresponding variational principle of the unconstrained min-max problem can be written as,
\be
\label{eq:8}
(\bfu^*,\bfphi^*)=\mathrm{arg \ \underset{\bfu}{min} \ \underset{\bfphi}{max}} \  \Pi(\bfu,\bfphi).
\ee
Here $\calW^{ext}$ is the work done by external sources. Denoting the boundaries of the domain by $\partial\Omega$ and edges by $C$, the work from external sources can be written as
\be
\label{eq:9}
\begin{aligned}
\calW&^{\Omega}=-b_iu_i+q\phi, \\ 
\calW&^{\partial\Omega}=-t_iu_i+r_i\delta^nu_i-w\phi, \\ 
\calW&^C=j_iu_i,
\end{aligned}
\ee
where $\bfb$ and $q$ are the external body force and electric charge density, $\bft$ and $\bfr$ are the traction and double traction on the boundary surfaces, $w$ is the surface charge density, and $\bfj$ is the surface tension, i.e., force per unit length. The symbol $\partial^n$ stands for normal derivative ${\partial / \partial\bfn}$. The domain boundary $\partial\Omega$ is split into Dirichlet and Neumann boundaries as
\be
\begin{aligned}
\label{eq:10}
\partial&\Omega=\partial\Omega_u \cup \partial\Omega_t = \partial\Omega_v \cup \partial\Omega_r = \partial\Omega_{\phi} \cup \partial\Omega_w, \text{and}\\
C&=C_u\cup C_j,
\end{aligned}
\ee
where $\partial\Omega_u$, $\partial\Omega_v$, and $\partial\Omega_\phi$ are the Dirichlet boundaries, with prescribed $\bfu$, its normal derivative $\bfv$, and $\bfphi$, respectively. $\partial\Omega_t$, $\partial\Omega_r$, and $\partial\Omega_w$ are the Neumann boundaries where traction $\bft$, double traction $\bfr$, and surface charge density $w$ are applied. The edges (corners in 2D) $C_u$ and $C_j$ correspond to the prescribed Dirichlet and Neumann boundary conditions in terms of displacement and surface tension $j$, respectively. The corresponding Dirichlet and Neumann boundary conditions are written as
\be
\label{eq:11}
\begin{split}
\begin{aligned}
    u&-\Bar{u}=0 \quad \text{on} \quad \partial\Omega_u,\\
    \partial&^nu-\Bar{v}=0 \quad \text{on} \quad \partial\Omega_v,\\
    \phi&-\Bar{\phi}=0 \quad \text{on} \quad \partial\Omega_\phi,\\
    u&-\Bar{u}=0 \quad \text{on} \quad C_u,
\end{aligned}
\end{split}
\quad \quad
\begin{split}
\begin{aligned}
    t&(u,\phi)-\Bar{t}=0 \quad \text{on} \quad \partial\Omega_t,\\
    r&(u,\phi)-\Bar{r}=0 \quad \text{on} \quad \partial\omega_r,\\
    w&(u,\phi)-\Bar{w}=0 \quad \text{on} \quad \partial\Omega_w,\\
    j&(u,\phi)-\Bar{j}=0 \quad \text{on} \quad C_j.
\end{aligned}
\end{split}
\ee

Using equations \eqref{eq:3}, and \eqref{eq:9} into \eqref{eq:7}, the total free enthalpy of the material can be written as
\begin{multline}
\label{eq:12}
\Pi [\bfu,\bfphi] = \int\limits_\Omega \left(\frac{1}{2}\dsC_{ijkl}\epsilon_{ij}\epsilon_{kl}+\frac{1}{2}h_{ijklmn}\epsilon_{ij,k}\epsilon_{lm,n}-\frac{1}{2}\kappa_{ij}E_iE_j-e_{ikl}E_i\epsilon_{kl}-\mu_{ijkl}E_i\epsilon_{jk,l}-b_iu_i \right. \\ \left. +q\phi\right) \mathrm{d\Omega}
- \int\limits_{\partial\Omega_t} t_iu_i \mathrm{d\Gamma} -
\int\limits_{\partial\Omega_r} r_i\partial^nu_i\mathrm{d\Gamma} + \int\limits_{\partial\Omega_w} w\phi \mathrm{d\Gamma} - \int\limits_{C_j} j_iu_i\mathrm{ds}.
\end{multline}
Here,
\begin{align}
\tag{13a}
\label{eq:13a}
t_i&\left(u,\phi\right) = \left(\hat{\sigma}_{ij}-\tilde{\sigma}_{ijk,k}- \nabla_k^S\tilde{\sigma}_{ikj}\right)n_j,\\
\tag{13b}
\label{eq:13b}
r_i&\left(u,\phi\right)= \tilde{\sigma}_{ijk}n_jn_k,\\
\tag{13c}
\label{eq:13c}
w & \left(u,\phi\right)=-\hat{D}_ln_l,\\
\tag{13d}
\label{eq:13d}
j_i&\left(u,\phi\right)=\left\llbracket\tilde{\sigma}_{ijk}m_jn_k\right\rrbracket,
\end{align}
where $\bfm$ is the outward pointing tangential vector on the concerned edges/corners, and $\llbracket \vdot \rrbracket$ is the jump operator, discussed further in the section \ref{subsec:2.3}. Following the formulation in \cite{codony2021mathematical}, and equation \eqref{eq:8} the variational principle at the saddle points $(\bfu^*,\bfphi^*)$ in enthalpy functional can be written as
\be
\tag{14}
\label{eq:14}
\delta\Pi(\bfu^*,\phi^*)=\delta\Pi_u(\bfu^*,\phi^*)+\delta\Pi_\phi(\bfu^*,\phi^*)=0, \qquad \forall (\delta \bfu,\delta \phi) \in \calU_0 \otimes \calP_0
\ee
with the functional spaces $\calU_0$, and $\calP_0$ defined as
\be
\tag{15}
\label{eq:15}
\begin{aligned}
\calU_0& = \{\delta \bfu \in [H^2(\Omega)]^D | \delta \bfu=0 \text{ on } \partial\Omega_u \text{ and } C_u, \partial^n \delta \bfu=0 \text{ on } \partial\Omega_v\}, \text{and} \\
\calP_0&=\{\delta\phi \in H^1(\Omega) | \delta\phi = 0 \text{ on } \partial\Omega_\phi \}.
\end{aligned}
\ee
Using equation \eqref{eq:12}, into equation \eqref{eq:14}, the weak form of the problem can be written as: \\

\textit{Find} $(\bfu,\phi)\in \calU_D \otimes \calP_D $ \textit{such that} $\forall (\delta \bfu,\delta\phi) \in \calU_0 \otimes \calP_0$
\begin{multline}
\tag{16}
\label{eq:16}
\int\limits_\Omega \left( \hat{\sigma}_{ij}\delta \epsilon_{ij} + \tilde{\sigma}_{ijk}\delta\epsilon_{ij,k} - \hat{D}_l\delta E_l\right)\mathrm{d\Omega} = \int\limits_\Omega \left(b_i\delta u_i-q\delta\phi\right)\mathrm{d\Omega} + \int\limits_{\partial\Omega_t} t_i\delta u_i \mathrm{d\Gamma} + \int\limits_{\partial\Omega_r}r_i\partial^n \delta u_i \mathrm{d\Gamma} \\ - \int\limits_{\partial\Omega_w} w\delta \phi \mathrm{d\Gamma} + \int\limits_{C_j}j_i\delta u_i \mathrm{ds},
\end{multline}
where functional spaces $\calU_D$, and $\calP_D$ are defined as
\be
\tag{17}
\label{eq:17}
\begin{aligned}
\calU_D&=\{\bfu \in [H^2(\Omega)]^D | \ \bfu=\Bar{\bfu} \ \text{on} \ \partial\Omega_u \ \text{and} \ C_u, \partial^n \bfu =\Bar{v} \ \text{on} \ \partial\Omega_v\}
, \text{and} \\
\calP_D&= \{\phi \in H^1(\Omega) | \ \phi=\Bar{\phi} \ \text{on} \ \partial\Omega_\phi\}.
\end{aligned} 
\ee
Integrating by parts and applying divergence theorem to equation \eqref{eq:16}, and using the definitions of state variables from equations \eqref{eq:4}-\eqref{eq:6}, the set of governing equations of the system can be written as
\be
\tag{18}
\label{eq:18}
\begin{aligned}
(\hat{\sigma}_{ij}&-\Tilde{\sigma}_{ijk,k})_{,j}+b_i=0 \ \ \text{in} \ \Omega, \\
\hat{D}_{l,l}&-q=0 \ \ \text{in} \ \Omega.
\end{aligned}
\ee
It is noteworthy that the solution of weak form in equation \eqref{eq:16} requires at least a $C^1$ continuous approximation and hence classical $C^0$ FE formulations can not be used. To resolve this issue we resort to IGA, which ensures the continuity of derivatives within the individual topological features (which can be modeled as a single NURBS patch), and a discontinuous Galerkin method is employed to ensure the continuity across different features (or patches).

\subsection{Discontinuous Galerkin method}
\label{subsec:2.3}
The NURBS basis of IGA provides continuous derivatives within the patches, while derivatives are discontinuous on the patch boundaries in a multi-patch domain. Considering the whole domain as a union of patches, i.e., $\{ \Omega_p\}_{p=1}^{np}$, the NURBS approximation will be in $\calH^2(\Omega_p)$,
\begin{multline}
\tag{19}
\label{eq:19}
\int\limits_{\Omega_p} \left( \hat{\sigma}_{ij}\delta \epsilon_{ij} +  \tilde{\sigma}_{ijk}\delta\epsilon_{ij,k} - \hat{D}_l\delta E_l  - b_i\delta u_i+q\delta\phi\right) \mathrm{d\Omega} - \int\limits_{\partial\Omega_{t_p}} t_i\delta u_i \mathrm{d\Gamma} - \int\limits_{\partial\Omega_{r_p}}r_i\partial^n \delta u_i \mathrm{d\Gamma} \\ + \int\limits_{\partial\Omega_{w_p}} w\delta \phi \mathrm{d\Gamma} - \int\limits_{C_{j_p}}j_i\delta u_i \mathrm{ds} = 0.
\end{multline}
Further, summing over patches and following the method in \cite{engel2002continuous,ventura2021c0},
\begin{multline}
\tag{20}
\label{eq:20}
\int\limits_{\Omega} \hat{\sigma}_{ij}\delta \epsilon_{ij}\mathrm{d\Omega} + \int\limits_{\hat{\Omega}} \tilde{\sigma}_{ijk}\delta\epsilon_{ij,k}\mathrm{d\Omega} - \int\limits_{\Omega} \left( \hat{D}_l\delta E_l - b_i\delta u_i + q\delta\phi\right) \mathrm{d\Omega} - \int\limits_{\calI} (t_i^L+t_i^R)\delta u_i\mathrm{ds} \\ - \int\limits_{\calI}\left(\left(\partial^n\delta  u_i\right)^L r_i^L+ \left (\partial^n\delta u_i\right)^R r_i^R\right)\mathrm{ds} - \int\limits_{\partial\Omega_{t}} t_i\delta u_i \mathrm{d\Gamma} - \int\limits_{\partial\Omega_{r}}r_i\partial^n \delta u_i \mathrm{d\Gamma} + \int\limits_{\partial\Omega_{w}} w\delta \phi \mathrm{d\Gamma} - \int\limits_{C_{j}}j_i\delta u_i \mathrm{ds} = 0.
\end{multline}
Here, $\hat{\Omega}$ represents the union of interiors of all patches, and $\calI$ is the union of internal patch boundaries, i.e.,
\begin{equation*}
\hat{\Omega}=\bigcup_{p=1}^{np}\Omega_e, \quad \mathrm{and} \quad \calI=\bigcup_{p=1}^{np}\partial\Omega_e \backslash \partial\Omega.
\end{equation*}
The superscripts $L$, and $R$ represent the values evaluated in the patch located at left and right of the boundary $\calI$, respectively. While continuity of the primary variables and equilibrium of forces on patch boundaries is satisfied during the assembly, the continuity of derivatives of primary variables and the equilibrium of higher-order traction is not fulfilled automatically and requires additional considerations. Defining jump and average operators as
\be
\tag{21}
\label{eq:21}
\llbracket a \rrbracket = a^L + a^R, \quad \mathrm{and} \quad
\{a\}=\frac{1}{2}\left(a^L+a^R\right);
\ee
the higher-order Dirichlet and Neumann boundary conditions can be written as
\be
\tag{22}
\label{eq:22}
\begin{aligned}
\llbracket& \partial^n\textbf{u} \rrbracket=0, \ \mathrm{and} \\
\llbracket& \textbf{r} \otimes \textbf{n} \rrbracket=0.
\end{aligned}
\ee
The integrand with the higher order traction in equation \eqref{eq:20} can be further simplified by using the relation $(a^L\textbf{n}^L)b^L+(a^R\textbf{n}^R)b^R=\{a\}\llbracket b\textbf{n} \rrbracket + \llbracket a\textbf{n} \rrbracket\{b\}$,
\be
\tag{23}
\label{eq:23}
\begin{aligned}
\left(\partial^n\delta  u_i\right)^L r_i^L+ \left (\partial^n\delta  u_i\right)^R r_i^R =& \{\partial^n \delta u_i\} \textbf{:} \llbracket r_i \otimes \textbf{n} \rrbracket + \llbracket \partial^n \delta u_i \rrbracket \vdot \{r_i\} \\
=& \llbracket \partial^n \delta u_i \rrbracket \vdot \{r_i\}.
\end{aligned}
\ee
Replacing equation \eqref{eq:23}, and equilibrium on the interior boundaries $\llbracket t_i \rrbracket = 0$ into equation \eqref{eq:20},
\begin{multline}
\tag{24}
\label{eq:24}
\int\limits_{\Omega} \hat{\sigma}_{ij}\delta \epsilon_{ij}\mathrm{d\Omega} + \int\limits_{\hat{\Omega}} \tilde{\sigma}_{ijk}\delta\epsilon_{ij,k}\mathrm{d\Omega} - \int\limits_{\Omega} \left( \hat{D}_l\delta E_l - b_i\delta u_i + q\delta\phi\right) \mathrm{d\Omega} \\ - \int\limits_{\calI}\llbracket \partial^n u_i \rrbracket \vdot \{r_i\}\mathrm{ds} - \int\limits_{\partial\Omega_{t}} t_i\delta u_i \mathrm{d\Gamma} - \int\limits_{\partial\Omega_{r}}r_i\partial^n \delta u_i \mathrm{d\Gamma} + \int\limits_{\partial\Omega_{w}} w\delta \phi \mathrm{d\Gamma} - \int\limits_{C_{j}}j_i\delta u_i \mathrm{ds} = 0.
\end{multline}
To ensure the stability of the solution, an interior penalty method-based stabilization is employed, yielding the weak form as
\textit{Find} $(u,\phi)\in \calU_D \otimes \calP_D $ \textit{such that} $\forall (\delta u,\delta\phi) \in \calU_0 \otimes \calP_0$
\begin{multline}
\tag{25}
\label{eq:25}
\int\limits_{\Omega} \hat{\sigma}_{ij}\delta \epsilon_{ij}\mathrm{d\Omega} + \int\limits_{\hat{\Omega}} \tilde{\sigma}_{ijk}\delta\epsilon_{ij,k}\mathrm{d\Omega} - \int\limits_{\Omega} \left( \hat{D}_l\delta E_l - b_i\delta u_i + q\delta\phi\right) \mathrm{d\Omega} - \int\limits_{\calI}\llbracket \partial^n \delta u_i \rrbracket \vdot \{r_i\}\mathrm{ds} -\int\limits_{\calI}  \{\delta r_i\} \llbracket \vdot \partial^n u_i \rrbracket \mathrm{ds} \\ + \int\limits_\calI \tau \llbracket \partial^n\delta u\rrbracket \vdot \llbracket \partial^nu \rrbracket \mathrm{ds}- \int\limits_{\partial\Omega_{t}} t_i\delta u_i \mathrm{d\Gamma} - \int\limits_{\partial\Omega_{r}}r_i\partial^n \delta u_i \mathrm{d\Gamma} + \int\limits_{\partial\Omega_{w}} w\delta \phi \mathrm{d\Gamma} - \int\limits_{C_{j}}j_i\delta u_i \mathrm{ds} = 0,
\end{multline}
where
\begin{equation*}
\begin{aligned}
\calU&_D=\{\bfu \in [H^1(\Omega) \cap H^2(\hat{\Omega})]^D | \ \bfu=\Bar{\bfu} \ \text{on} \ \partial\Omega_u \ \text{and} \ C_u, \partial^n \bfu =\Bar{v} \ \text{on} \ \partial\Omega_v\}, \\ 
\calV&_D=\{\phi \in H^1(\Omega)| \ \phi=\Bar{\phi} \ \text{on} \ \partial\Omega_\phi\},\\
\calU&_0 = \{\delta \bfu \in [H^1(\Omega) \cap H^2(\hat{\Omega})]^D | \delta \bfu=0 \text{ on } \partial\Omega_u \text{ and } C_u, \partial^n \delta \bfu=0 \text{ on } \partial\Omega_v\}, \text{and} \\ 
\calP&_0=\{\delta\phi \in H^1(\Omega) | \delta\phi = 0 \text{ on } \partial\Omega_\phi \}.
\end{aligned}
\end{equation*}
Here $\tau$ is the penalty or stabilization parameter, which can either be calculated by solving an eigen-value problem or can directly be evaluated as \cite{engel2002continuous,ventura2021c0}
\be
\tag{26}
\label{eq:26}
\tau=\frac{\alpha E\calL^2}{h},
\ee
where $E$ is Young's modulus, $\calL$ is the length scale parameter, $h$ is the element size, and $\alpha$ is a constant whose value is tuned until a sufficient level of reduction in jump in the gradient is achieved at the patch interfaces.
\section{Isogeometric analysis}
\label{sec:IGA}
\subsection{Discretization}
\begin{figure}[ht!]
    \centering
    \includegraphics[width=\textwidth]{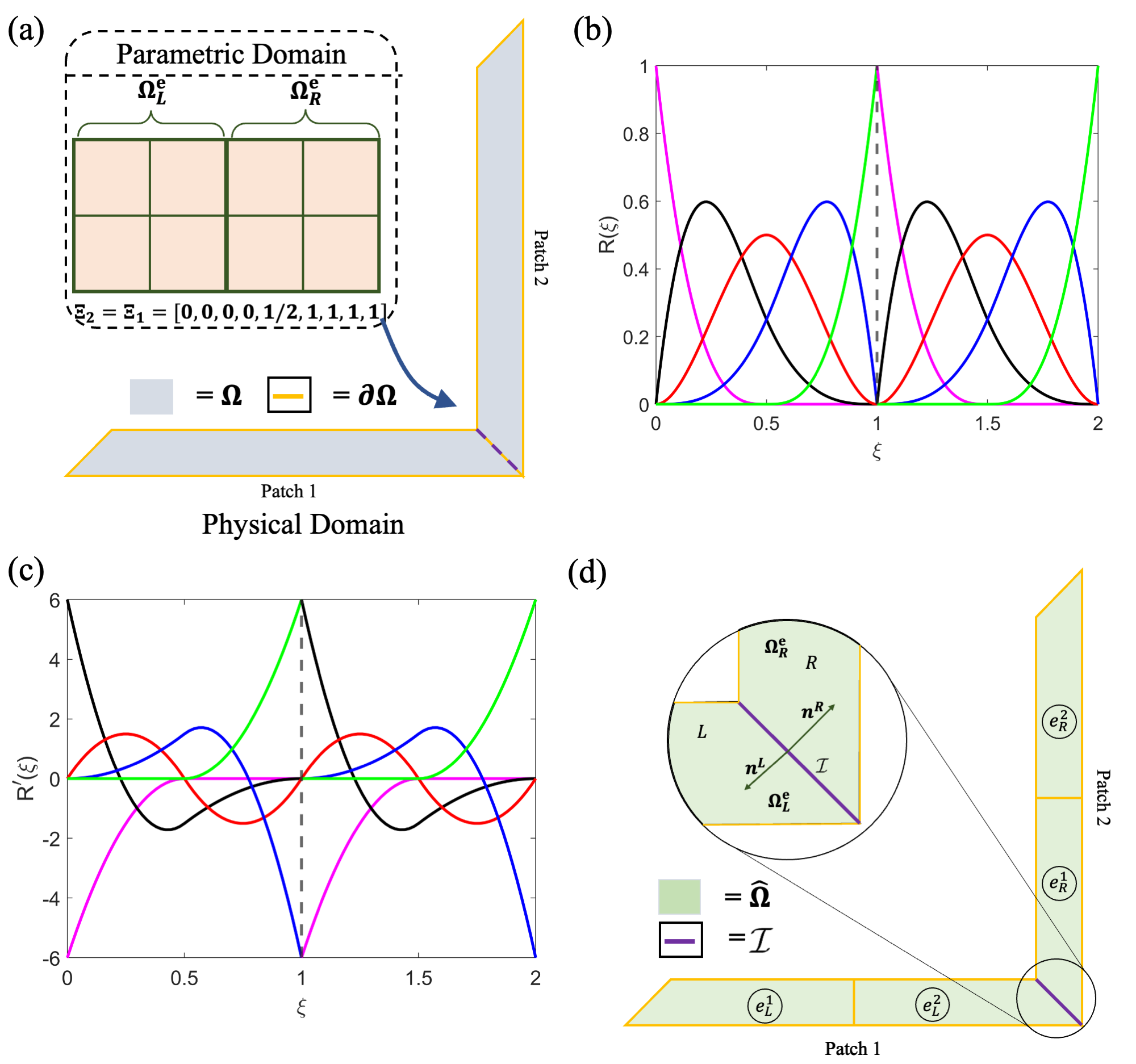}
    \caption{Schematic of weak enforcement of $C^1$ continuity in a two-patch domain IGA. \textbf{(a)} The physical domains of an L-shaped beam modeled with two patches whose boundary is highlighted as a dashed line. The mapping of the parametric domain to the physical domain, defined by the knot vectors $\Xi_1=\Xi_2=\left[0, 0, 0, 0, 1/2, 1, 1, 1, 1\right]$, is shown in the inset. \textbf{(b)} NURBS basis function for the two patches, exhibiting $C^0$ continuity at the boundary, and \textbf{(c)} the derivatives of NURBS basis functions, which are discontinuous at the patch boundaries. \textbf{(d)} Representation of different domains, boundaries, and normal vectors used in the implementation of interior penalty-based discontinuous Galerkin method.}
    \label{fig:IGA}
\end{figure}
Isogeometric analysis (IGA) -- originally developed for integrating CAD and CAE -- proves to be advantageous in solving fourth-order partial differential equations, due to its NURBS basis. NURBS basis functions have up to $(p-1)$ continuous derivatives across element boundaries, where $p$ is the polynomial order. The geometry of the domain is approximated through NURBS basis functions and control points' coordinates as
\be
\tag{27}
\label{eq:27}
(x,y)=X(\xi,\eta)=\sum\limits_{i=1}^{n_x}\sum\limits_{j=1}^{n_y} R_{i,j}^{p,q}(\xi,\eta)\tilde{X}_{ij},
\ee
where, $n_x$, and $n_y$ are the number of control points in $x$ and $y$ directions, and $p_x$ and $p_y$ are the orders of polynomial in parametric directions $\xi$ and $\eta$. The NURBS functions of order $p$ and $q$ in parametric directions $\xi$ and $\eta$, at a point with indices $i$, $j$  are defined in terms of B-spline basis functions $N_i(\xi)$ as
\begin{equation*}
    R_{i,j}^{p,q}(\xi,\eta)=\frac{N_{i,p}(\xi)M_{j,q}(\eta)w_{i,j}}{\sum\limits_{\hat{i}=1}^n \sum\limits_{\hat{j}=1}^m N_{\hat{i},p}(\xi) M_{\hat{j},q}(\eta)w_{\hat{i},\hat{j}}},
\end{equation*}
where $w$ denotes the weights associated with the point of evaluation. 
The same approximation is used to evaluate the degrees of freedom, i.e., displacements and electric potential as follows,
\be
\tag{28}
\label{eq:28}
\begin{aligned}
u&(\xi,\eta)=\sum\limits_{i=1}^{n_x}\sum\limits_{j=1}^{n_y} R_{i,j}^{p,q}(\xi,\eta)\tilde{u}_{ij}, \ \text{and} \\
\phi&(\xi,\eta)=\sum\limits_{i=1}^{n_x}\sum\limits_{j=1}^{n_y} R_{i,j}^{p,q}(\xi,\eta)\tilde{\phi}_{ij}.
\end{aligned}
\ee
Here $u=\{u_1,u_2\}$ is the displacement field containing displacements along $x$, and $y$ directions. Superscript $^*$ denotes the values at control points of the respective variable. The mechanical strain, electric field, and their gradients can be written in the form of derivative matrices as

\be
\tag{29}
\label{eq:29}
\begin{aligned}
\bm{\epsilon}&=\bfB_u^T\tilde{\bfu}, \quad \nabla\bm{\epsilon}=\bfH_u^T\tilde{\bfu}, \\
\bfE&=-\bfB_\phi^T\tilde{\bm{\phi}}, \quad \nabla\bfE=-\bfH_\phi^T\tilde{\bm{\phi}},
\end{aligned}
\ee
where $\bfepsilon$ and $\bfE$ are the strain and electric field vectors; and $\bfB_u$, $\bfB_\phi$, $\bfH_u$, and $\bfH_\phi$ are the matrices containing the first and second derivatives of the NURBS basis functions. Expanded forms of these vectors and matrices are provided in \ref{sec:appendixA}.

\subsection{Discretized system of equations} 
The weak form in the equation \eqref{eq:25} can be expressed in matrix form as
\begin{multline}
\tag{30}
\label{eq:30}
\Assembly\limits_{n=1}^{n_{el}}\left[\int\limits_{\Omega^e} \left(\delta \bm{\epsilon}^T \hat{\bm{\sigma}} - \delta\bfE^T\hat{\bfD}\right)\mathrm{d\Omega} +\int\limits_{\hat{\Omega}^e} \delta\nabla\bm{\epsilon}^T 
\tilde{\bm{\sigma}}\mathrm{d\Omega}\right] - \Assembly\limits_{I=1}^{n_I}\left[ \int\limits_{\calI^I} \left( \left\llbracket \frac{\partial\delta\bfu}{\partial\bfn} \right\rrbracket \vdot \{\bfr\} - \{\delta\bfr\} \vdot \left\llbracket \frac{\partial\bfu}{\partial\bfn} \right\rrbracket \right. \right. \\ \left. \left. +  \tau \left\llbracket \frac{\partial\delta\bfu}{\partial\bfn} \right\rrbracket \vdot \left\llbracket \frac{\partial\bfu}{\partial\bfn} \right\rrbracket \right) \mathrm{ds}\right]  =  \Assembly\limits_{n=1}^{n_{el}}\left[\int\limits_{\Omega^e} \left(\delta\bfu^T\bfb - \delta\bm{\phi}^T\bfq\right) \mathrm{d\Omega} +\int\limits_{\partial\Omega_{t}^e} \delta\bfu^T\bft \mathrm{d\Gamma} + \int\limits_{\partial\Omega_{r}^e}\delta\left(\frac{\partial\bfu}{\partial\bfn}\right)^T\bfr  \mathrm{d\Gamma} \right. \\ \left. - \int\limits_{\partial\Omega_{w}^e} \delta \bm{\phi}^T\bfw \mathrm{d\Gamma} + \int\limits_{C_{j}^e}\delta\bfu^T\bfj \mathrm{ds}\right],
\end{multline}
where $n_{el}$ and $n_I$ denote the total number of elements and patch boundaries in the domain, and A is the assembly operator. The last term associated with the corners $C_j^e$ reduces to a sum of forces on the corner, contributed by all the elements sharing that corner. This term is included for writing a well-defined boundary value problem, and it is zero on all the interior corners due to the internal equilibrium of the forces, while on the corners of exteriors, it is equal to the prescribed point forces (line force in 3D). Further, equation \ref{eq:30} can be expressed in compact form as
\be
\tag{31}
\label{eq:31}
\left(\bfK + \bfK_\calI\right)\bfU = \bfF,
\ee
where
\begin{equation*}
\begin{aligned}
\bfK&=
    \begin{bmatrix}
        \bfK_{uu} & \bfK_{u\phi}\\
        \bfK_{\phi u} & -\bfK_{\phi\phi}
    \end{bmatrix}, \\
\bfF&=
\begin{bmatrix}
\bfF_U\\
\bfF_{\phi}
\end{bmatrix}, \quad 
\bfU=
\begin{bmatrix}
\tilde{\bfu}\\
\tilde{\bm{\phi}}
\end{bmatrix}.
\end{aligned}
\end{equation*}
The stiffness and force matrices are defined using equations \eqref{eq:27}-\eqref{eq:29} as
\be
\tag{32a}
\label{eq:32a}
\bfK_{uu} = \Assembly\limits_{n=1}^{n_{el}}\int\limits_{\Omega^e}\left(\bfB_u\bfC\bfB_u^T + \bfH_u\bfh\bfH_u^T\right) \mathrm{d\Omega},
\ee
\be
\tag{32b}
\label{eq:32b}
\bfK_{u\phi}=\bfK_{\phi u}^T=\Assembly\limits_{n=1}^{n_{el}}\int\limits_{\Omega^e}\left(\bfB_{\phi}\bfe\bfB_{u}^T + \bfB_{\phi}\bm{\mu}\bfH_u^T\right) \mathrm{d\Omega},
\ee
\be
\tag{32c}
\label{eq:32c}
\bfK_{\phi\phi}=\Assembly\limits_{n=1}^{n_{el}}\int\limits_{\Omega^e}\bfB_{\phi}\bm{\kappa}\bfB_{\phi}^T \mathrm{d\Omega},
\ee
\be
\tag{32d}
\label{eq:32d}
\bfF_u=\Assembly\limits_{n=1}^{n_{el}}\left(\int\limits_{\Omega}\bfR^T\bfb \mathrm{d\Omega} + \int\limits_{\partial\Omega_t^e}\bfR^T\bft\mathrm{d\Gamma} \right),
\ee
\be
\tag{32e}
\label{eq:32e}
\bfF_\phi=\Assembly\limits_{n=1}^{n_{el}}\left(\int\limits_{\Omega}\bfR^T\bfq \mathrm{d\Omega} + \int\limits_{\partial\Omega_w^e}\bfR^T\bfw \mathrm{d\Gamma} \right),
\ee
\be
\tag{32f}
\label{eq:32f}
\bfK_\calI= \Assembly\limits_{I=1}^{N_i}\int\limits_{\calI^I} \left( \left \llbracket \frac{\partial R_i}{\partial n} \right \rrbracket \vdot \left\{ \hat{\bfr}\right\} - \left\{\hat{\bfr}\right \} \vdot \left \llbracket \frac{\partial R_i}{\partial n} \right \rrbracket + \tau \left \llbracket \frac{\partial R_i}{\partial n} \right \rrbracket \vdot \left \llbracket \frac{\partial R_j}{\partial n} \right \rrbracket \right) \mathrm{ds},
\ee
where
\begin{equation*}
    \hat{\bfr} = \left(h_{ijklmn}R_{l,m,n} - \mu_{ijkl}R_{,l} \right) n_j n_k.
\end{equation*}

\subsection{Implementation in standard IGA workflow}
Due to the additional stiffness contribution at the interior boundaries of patches, the implementation of interior penalty-based DG method requires incorporating some extra steps in a standard IGA code. First, the assembly of mechanical, electromechanical, and electrical stiffness matrices (equations \eqref{eq:32a} - \eqref{eq:32c}) is performed as usual. To implement the DG method, first, all the interior patch boundaries (surfaces in 3D and edges in 2D) are identified and stored in an array. An assembly loop is run over the edges to compute the jumps and averages of the derivatives of the basis functions $R_{i,j}^{p,q}$. This involves the computation of first and second derivatives of $R_{i,j}^{p,q}$ on the two adjacent boundary elements of contiguous patches. The line integral (in 2D) $\bfK_\calI$ is computed over the Gauss points of edges shared by two or more patches. The value of stabilization or penalty parameter $\tau$ is taken large enough to ensure the smoothness of second derivatives across patches while keeping it reasonably finite to avoid ill-conditioning of the stiffness matrix. Since second derivatives are essential to compute the Hessian matrix, $p,q\geq2$ is necessary while computing NURBS basis functions. Here, we use $p=q=2$ to approximate both the displacement and the electric potential fields. Bezier extraction \cite{borden2011isogeometric} is employed to avoid recomputing basis functions and their derivatives in each iteration of the assembly loop, making assembly computationally efficient. Special consideration is required to maintain the conditioning of stiffness matrix in electromechanical problems, due to the huge difference in orders of magnitude of Young's modulus and dielectric permittivity, and hence diagonal elements of concatenated stiffness matrix $\bfK$. This is done by using a scaling parameter ($\beta\approx10^{10}$) on the stiffness matrix, and electrical degrees of freedom (both in the solution and while applying Dirichlet boundary conditions) as

\be
\tag{33}
\label{eq:33}
\bfK=
    \begin{bmatrix}
        \bfK_{uu} & \beta\bfK_{u\phi}\\
        \beta\bfK_{\phi u} & -\beta^2\bfK_{\phi\phi}
    \end{bmatrix}, \quad
\bfU=
\begin{bmatrix}
\tilde{\bfu}\\
\frac{1}{\beta}\tilde{\bm{\phi}}
\end{bmatrix}.
\ee
It is worth mentioning that the contribution of interior patch boundaries to the overall stiffness matrix (equation \eqref{eq:32f}) is not computed in the matrix form. The reason behind this is the complexity of vectorization of the scattered boundary contributions, which are very difficult to write in the form of separate matrices of basis functions and their derivatives. Owing to the low number of interior patch boundaries, thanks to the higher order intrapatch continuity of NURBS basis, computing $\bfK_\calI$ in a separate loop does not make much difference computationally.
\section{Validation of the multipatch coupling}
\label{sec:Validation}
\subsection{Continuity of the gradients}
\begin{figure}[ht]
    \centering
    \includegraphics[width=\textwidth]{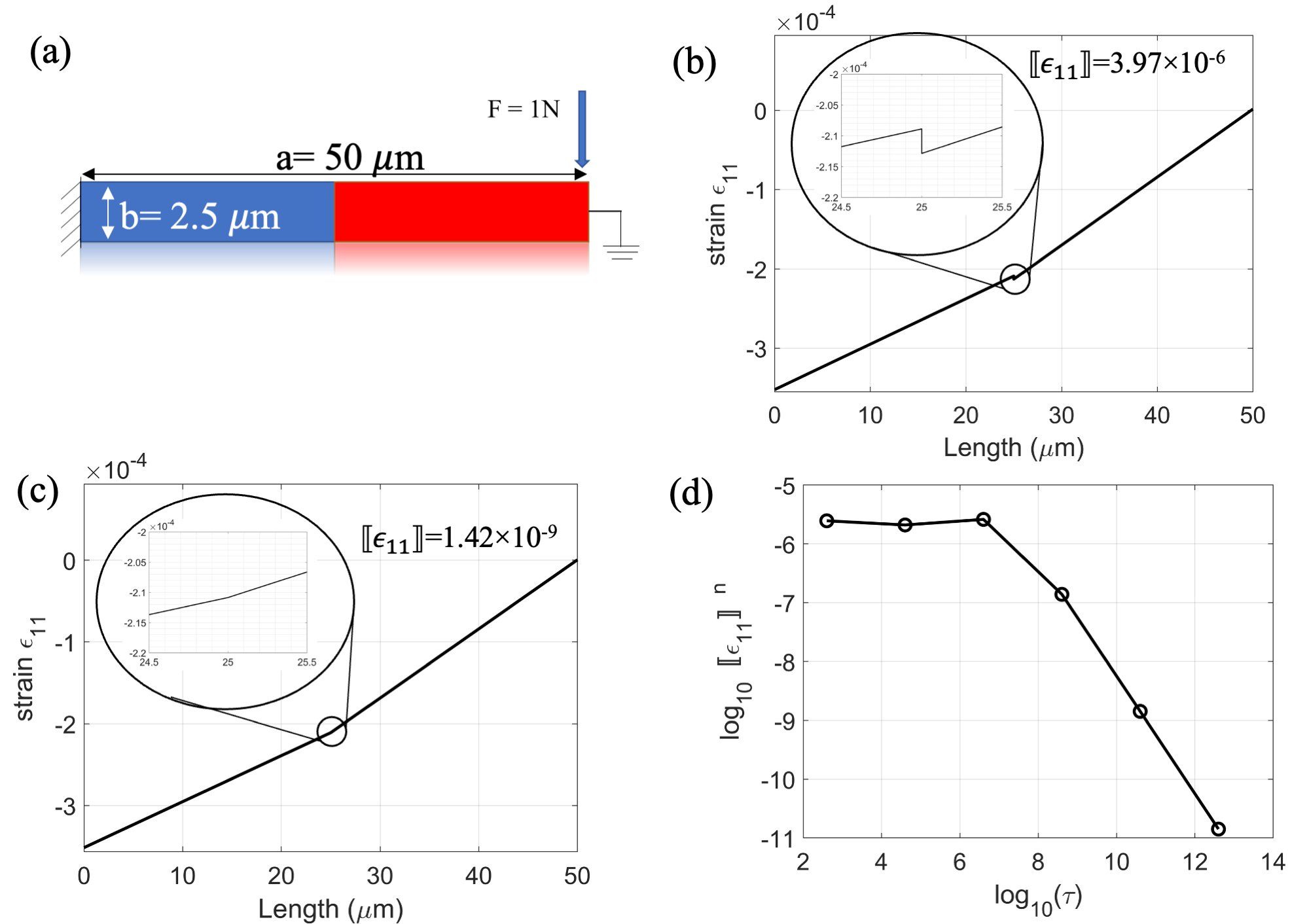}
    \caption{Weak enforcement of continuity of gradients in a two-patch cantilever example. \textbf{(a)} The geometry and boundary conditions of the cantilever beam, discretized with two patches, \textbf{(b)} variation of axial strain $\epsilon_{11}$ along the length of the beam computed with usual $C^0$ continuous patch interface, \textbf{(c)} variation of $\epsilon_{11}$ with continuity of gradients enforced by interior penalty based DG method with a stabilization parameter $\tau=10^{10}$, and \textbf{(d)} variation of normalized jump with increasing value of $\tau$ on a logarithmic scale, where $\llbracket\epsilon_{11}\rrbracket^n = \llbracket \epsilon_{11} \rrbracket / \epsilon_{11}^{\mathrm{max}}$}
    \label{fig:cont}
\end{figure}

The enforcement of the continuity of gradients of the primary variables through the DG method is demonstrated in this section. A cantilever beam under open-circuit boundary conditions is subjected to a unit point load at the free end, as shown in Fig \ref{fig:cont}(a). The material parameters of the beam are given in Table \ref{tab:cant_para}. The beam is discretized with two patches whose interface would have discontinuous derivatives of the primary variables, and hence a jump in the axial strain, $\epsilon_{11}$ in standard IGA settings. Note that each patch is discretized with only one element, i.e., no knot refinement is performed since the focus here is only to demonstrate the elimination of discontinuity across patch interfaces. The order of basis function in parametric directions $\xi$, and $\eta$ is taken as $p=q=3$ for all the studies in this work. Fig \ref{fig:cont}(b) shows the variation of $\epsilon_{11}$ along the length of the beam, computed using the $C^0$ continuous multipatch weak form given by equation (\ref{eq:16}). A \textit{kink} at the interface of the two patches can be observed in the plot. The implementation of weak form in equation (\ref{eq:25}) yields a smooth variation of strain along the length, reducing jump to negligible magnitudes. Fig. \ref{fig:cont} (c) shows the variation of $\epsilon_{11}$ computed by solving IGA system of equations (\ref{eq:31}) with stabilization parameter $\tau=4 \times 10^{10}$, demonstrating a reduction by a factor of $\approx10^3$. Further, the variation of the relative jump in axial strain $\epsilon_{11}$ with $\tau$ is plotted on a logarithmic scale in Fig \ref{fig:cont}(d). A continuous reduction in relative error can be observed with increasing values of $\tau$, however, one needs to be mindful of maintaining the conditioning of the stiffness matrix while increasing the value of $\tau$.
\begin{table}[ht!]
    \caption{Material parameters used in the simulation studies.}
    \centering
    \begin{tabular}{|c | c | c | c |c |c |c|}
    \hline
           E (Pa) & $\nu$ & $\mu_{11}$ (C/m) & $\mu_{12}$ (C/m) & $e_{11}$ (C/m) & $e_{21}$ (C/m) & $\kappa_{11}=\kappa_{22}$ (C/Vm) \\
        \hline
        $100\times 10^9$ & 0.37 & $1\times 10^{-6}$ & $1\times 10^{-6}$ & 4.4 & -4.4 & $12.48\times10^{-9}$ \\
        \hline
    \end{tabular}
    \label{tab:cant_para}
\end{table}

\begin{figure}[ht]
    \centering
    \includegraphics[width=\textwidth]{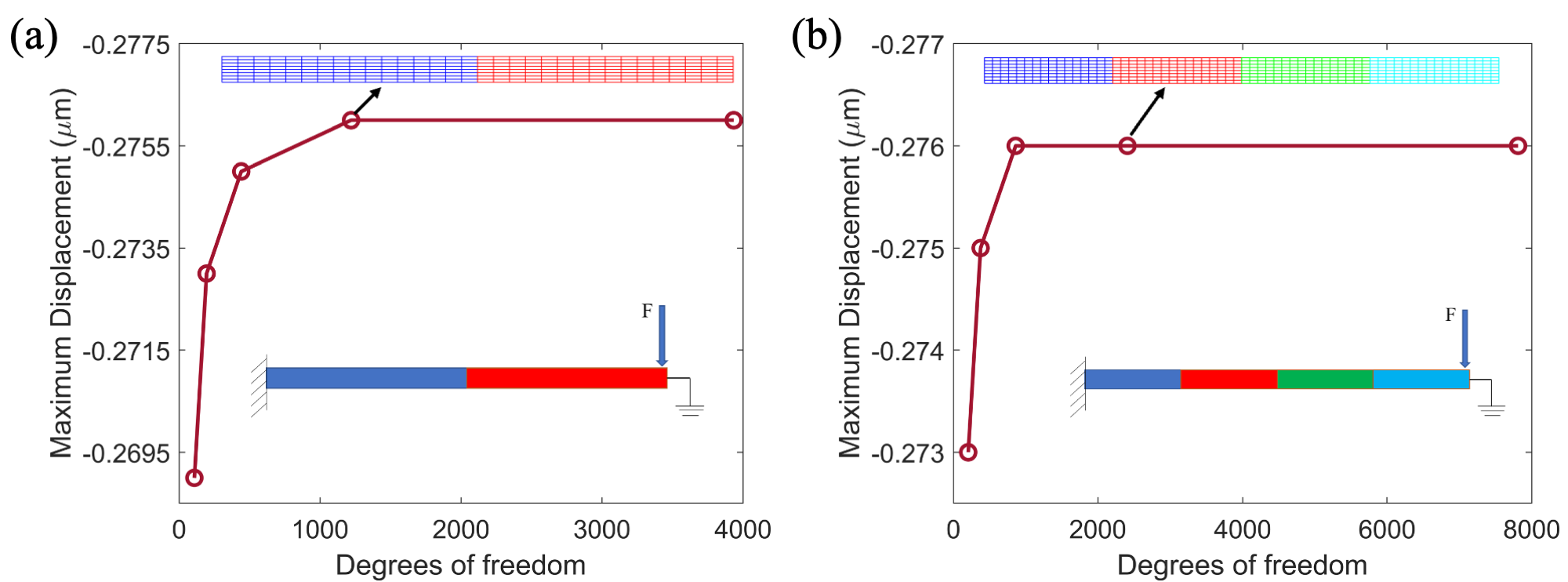}
    \caption{Mesh convergence study for a cantilever beam discretized with \textbf{(a)} two, and \textbf{(b)} four patches, with converged meshes for both the cases shown in the inset.}
    \label{fig:conv_beam}
\end{figure}

Further, to showcase the robustness of $C^1$ continuous patch interfaces, the same example of a cantilever beam under open circuit boundary conditions is checked for convergence. The beam is modeled with two and four patches and the convergence of maximum displacement with the number of degrees of freedom is plotted. Fig \ref{fig:conv_beam}(a) and (b) show an excellent convergence observed for both, two and four-patch cases.
\subsection{Verification of the computational model}
\begin{figure}[ht!]
    \centering
    \includegraphics[width=\textwidth]{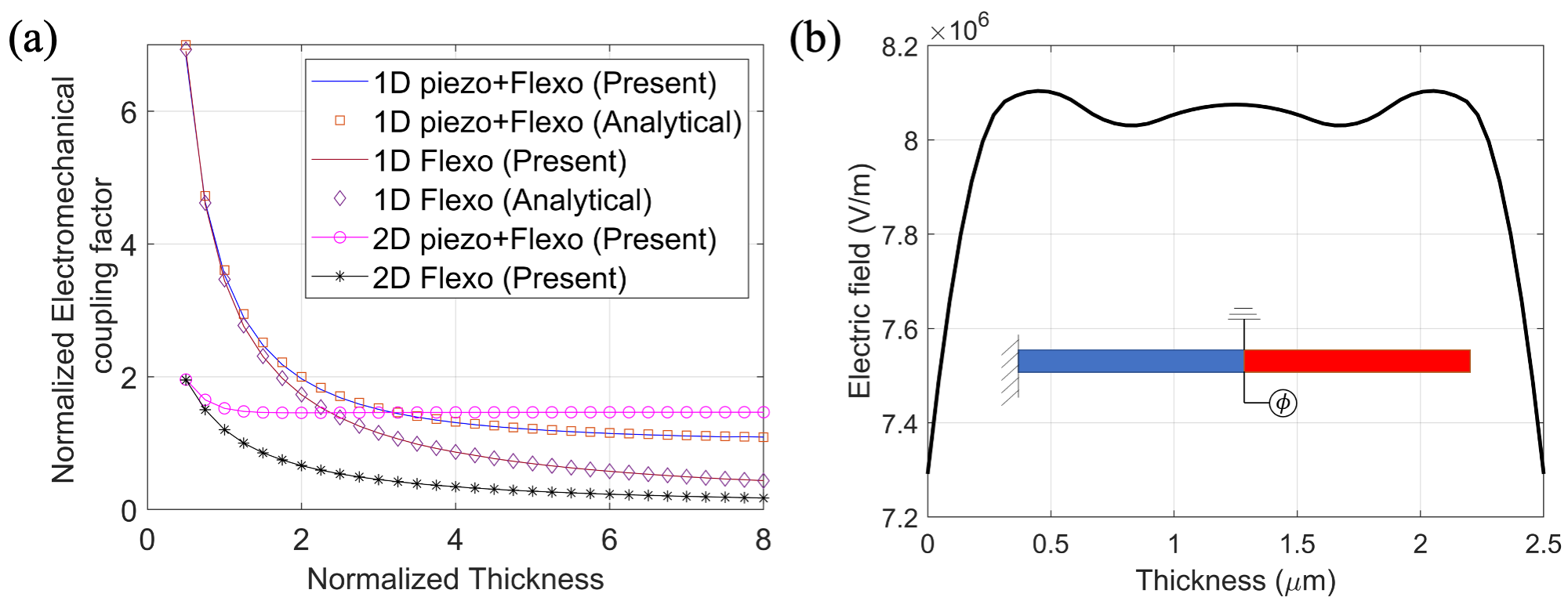}
    \caption{Validation of the present model against analytical calculations of electromechanical coupling factor. \textbf{(a)} Variation of electromechanical coupling factor with normalized thickness under unit load and open circuit boundary condition, showing great match with the results for pure flexoelectric and piezoelectric-flexoelectric couplings. \textbf{(b)} variation of electric field across the thickness of the beam, under an applied voltage $\phi=20 \mathrm{V}$ and closed circuit boundary conditions.}
    \label{fig:validation}
\end{figure}

Before proceeding to analyze the electromechanical behavior of architected dielectrics, we validate our model against the analytical results of Abdollahi et al. \cite{abdollahi2014computational} for homogeneous dielectric solids. The normalized electromechanical coupling factor -- defined as the ratio of total electrical and mechanical energies -- of a cantilever beam is evaluated for different combinations of piezoelectric and flexoelectric couplings under open circuit conditions and application of unit load, as shown in Fig \ref{fig:validation}(a). Both 2D and 1D cases are analyzed and compared with the analytical results. The electromechanical coupling factor $K_{EM}$ is defined as the ratio of electrical energy generated and the input mechanical energy. For a cantilever beam it can be expressed as \cite{abdollahi2014computational,sharma2021performance} 
\be
\label{eq:34}
K_{EM}^{Beam}=\frac{\chi}{1+\chi}\sqrt{\frac{\kappa}{Y}\left(e^2+12\left(\frac{\mu}{t}\right)\right)},
\ee
where $\chi$ is the electric susceptibility, given as $\chi=\kappa+1$, $Y$ is the Yonug's modulus, and $t$ is the thickness of the beam. For 1D cases, $\kappa=\kappa_{22}$, $e=e_{21}$, and $\mu=\mu_{12}$ are considered in the analytical formula, while in computational model, $\kappa_{11}$, $\mu_{11}$, $e_{11}$, and $\nu$ are set to zero to simulate 1D cases. The normalized values of $K_{EM}^{Beam}$ are calculated by dividing it with $K_{EM}^{Beam}$ for pure piezoelectric material, while normalized thickness is defined as $h^\prime=-eh/\mu$. To test the validity of the present DG method-based IGA, the validation studies are conducted for a two-patch discretization. An excellent agreement of present results with the analytical calculations is observed. Further, the present model is tested under closed circuit boundary conditions, by applying an electric potential of 20 V at the bottom surface and grounding the top surface. The variation of electric field across the thickness properly captures the effect of flexoelectricity on the electric field distribution, as the sharp gradients near the top and bottom surfaces can be observed.

\section{Electromechanical behavior of 2D dielectric trusses}
\label{sec:Results}
Among the various classes of architected materials that have emerged in recent years, truss-based architected materials or \textit{metamaterials} remain the most common choice. The motivation for employing architected materials is to achieve tailored and superior physical properties per unit volume, facilitated by the design of the microstructure rather than mere chemical composition. In this study, we analyze the potential of truss lattices to realize flexoelectricity under uniform macro-strains and boundary conditions in \textit{non-piezoelectric} materials, i.e., hereafter contribution of piezoelectricity is omitted in this study. Due to the strain-gradient dependence of flexoelectricity, having microstructural features enables electromechanical coupling in general dielectric materials. Using the $C^1$ continuous multi-patch IGA framework, we simulate 2D truss structures with different combinations of boundary conditions and architectures. Specifically, we choose four \textit{unit cells} (UCs) for constructing these truss structures as shown in Fig \ref{fig:UCs-compression}.

\subsection{Direct flexoelectric effect}
\begin{figure}
    \centering
    \includegraphics[width=\textwidth,height=\dimexpr \textheight - 7\baselineskip\relax, keepaspectratio]{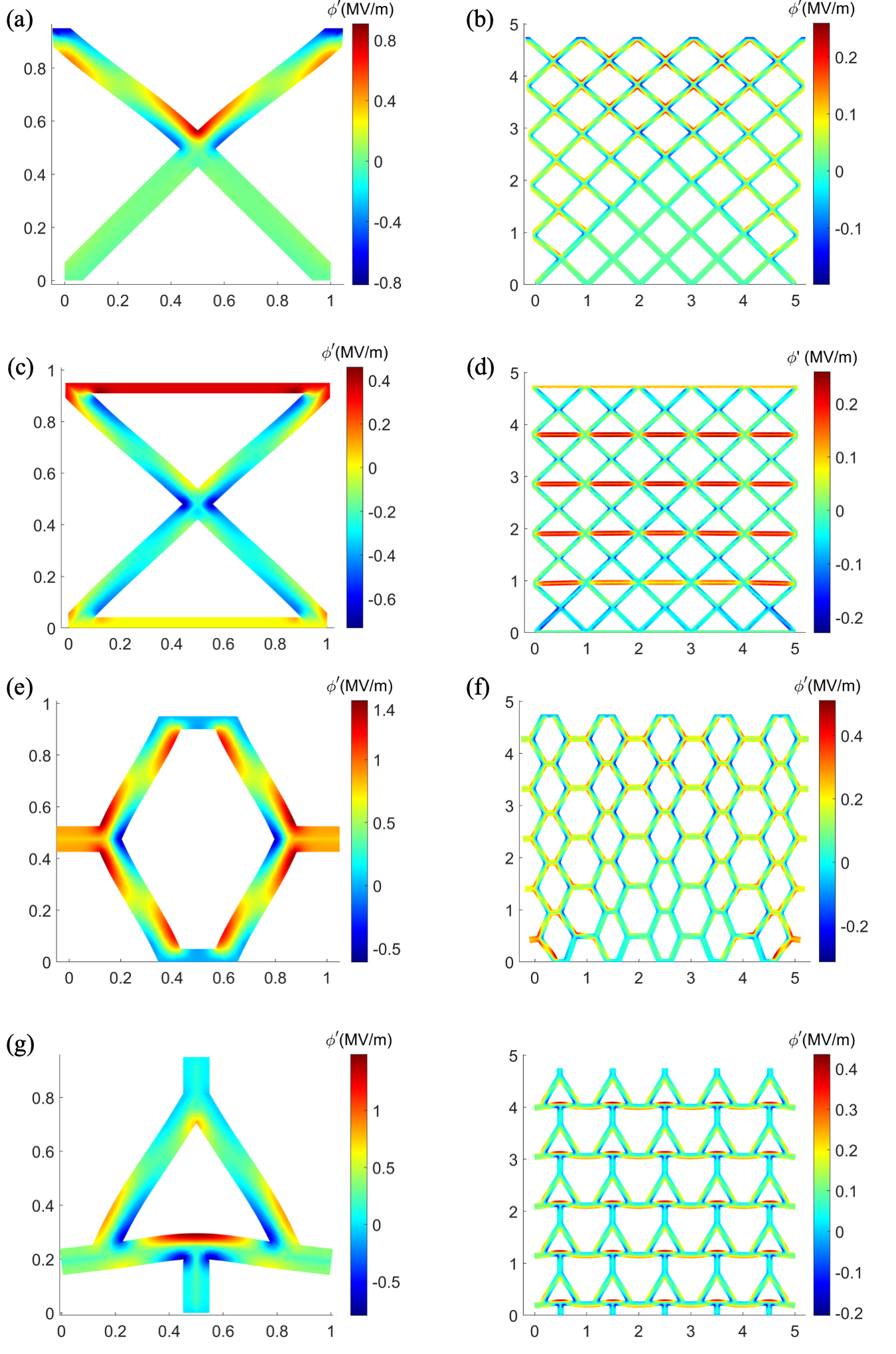}
    \captionsetup{justification=justified}
    \caption{Normalized electric potential $(\phi^\prime=\phi/b \ \mathrm{V/m})$ distribution in the truss lattices based on four selected unit cells under compressive load and open circuit boundary conditions. Single unit cells under compression, termed as UC1 - UC4 (\textbf{a}, \textbf{c}, \textbf{e}, and \textbf{f}, respectively). All the UCs are electrically ground at the bottom and the top surface is subjected to equipotential electrical constraint.  High magnitudes of electric potential at the corners can be observed due to high strain gradients. \textbf{(b,d,f,h)} show the normalized electric potential distribution in $5\times5$ tessellations of the UC1-UC4. Due to strain gradients at the intersections of the struts, the electric potential is sustained at the larger sizes.}
    \label{fig:UCs-compression}
\end{figure}

To analyze the truss lattices in direct flexoelectricity or sensing mode, the four UCs are considered under compression, with width, and height $a=b=1 \mu m$, and relative density or fill fraction $\rho=0.2$. A compression equal to $b/20$, and equipotential boundary conditions are applied at the top surface, while all the degrees of freedom (mechanical and electrical) are set to zero at the bottom surface. The material properties used for simulations are as given in Table \ref{tab:cant_para}. Mesh convergence study is performed for for all the UCs considered here, as shown in \ref{sec:appendixB} (Fig \ref{fig:UC_convergence}). Note that since periodic boundary conditions are not applied here, the strain distribution is not uniform and hence even the centrosymmetric UCs (UC1-UC3) exhibit significant electric potential generation. The comparison of electric potential distribution in UC1 under symmetric and non-symmetric loading is shown in \ref{sec:appendixB} (Fig \ref{fig:Symmetric_BC}). Table \ref{tab:phi_compress} shows the normalized potential difference and electromechanical coupling factors for 1x and 5x tessellations of the four UCs, and the normalized electric potential distribution is plotted in Fig \ref{fig:UCs-compression}. In the case of a solid square-shaped specimen the generated electric potential and $K_{EM}$ are zero under compression, due to the uniform strain, however truss lattices facilitate non-uniform strain due to their structure; and thus do not require varying material composition or width to realize flexoelectricity. From Table \ref{tab:phi_compress}, it is observed that while the electromechanical coupling factor changes slightly from single UC to 5x tessellation, the normalized potential difference drops significantly due to the size effect of flexoelectricity. Notably, UC1 and UC4 show the highest mechanical-to-electrical energy conversion, due to bending dominated nature of deformation. Whereas, UC2 and UC3 exhibit relatively higher electric potential differences between top and bottom surfaces, which can be attributed to the higher electrode area available in these lattices.
\begin{table}
    \caption{Potential difference and electromechanical coupling factors in different truss lattices under compression.}
    \label{tab:phi_compress}
    \centering
    \small
    \resizebox{\textwidth}{!}{\begin{tabular}{|c | c | c | c |c |c |c |c |c|}
    \hline
           & UC1 (1x) & UC1 (5x) & UC2 (1x) & UC2 (5x) & UC3 (1x) & UC3 (5x) & UC4 (1x) & UC4 (5x)\\
        \hline
        $\Delta\phi^\prime (\mathrm{MV/m})$ & $0.681$ & $0.123$ & $0.349$ & 0.087 & 0.070 & 0.060 & 0.467 & 0.025\\
        \hline
        $K_{EM}$ & 0.212 & 0.211 & 0.081 & 0.149 & 0.189 & 0.182 & 0.228 & 0.206\\
        \hline
    \end{tabular}}
\end{table}

\begin{figure}
    \centering
    \includegraphics[width=\textwidth, height=\dimexpr \textheight - 5\baselineskip\relax, keepaspectratio]{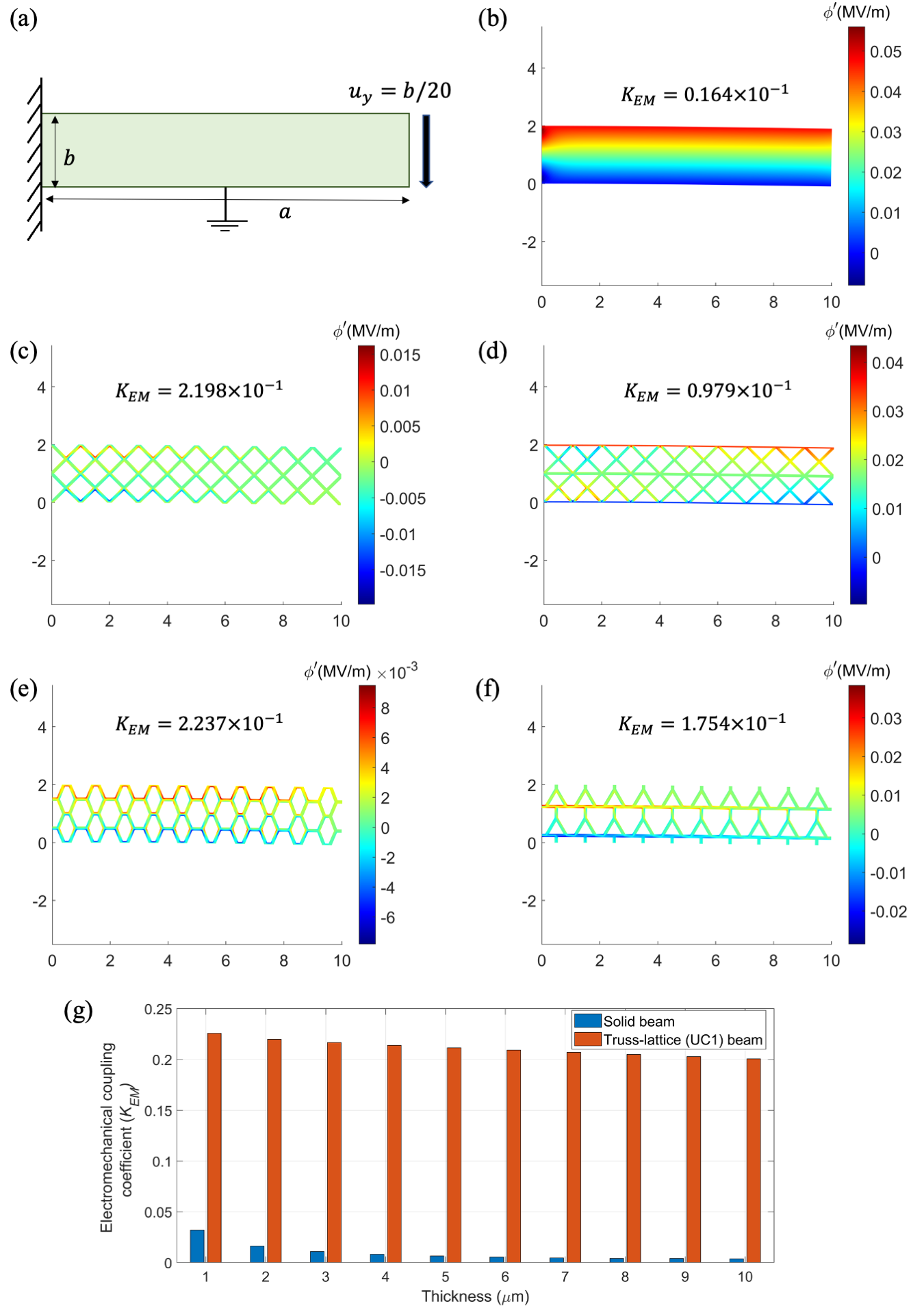}
    \caption{Electric potential distribution in truss lattices in bending mode. \textbf{(a)} Geometry and boundary conditions used for the analyses in (b)-(f), \textbf{(b)} normalized electric potential ($\phi^\prime$) distribution in a solid cantilever beam, and  \textbf{(c-f)}  electric potential distribution in beams of truss lattices constructed by tessellating UC1 - UC4. \textbf{(g)} Comparison of $K_{EM}$ in solid and UC1-based truss lattices, showing the ability to maintain higher electrical to mechanical energy conversion in truss lattices.}
    \label{fig:Bending-direct}
\end{figure}

Next, we consider the truss lattices under bending mode. The cantilever beam of each design is considered with a $10\times 2$ tessellation, i.e., of dimensions $10\times 2 \ \mu m$, and a relative density of 0.2. The beam is subjected to a vertical deflection of $b/20$ m at the free end, while the bottom surface is electrically ground and the top surface is subjected to an equipotential constraint. Fig \ref{fig:Bending-direct} shows the electric potential distribution and electromechanical coupling factor of a solid beam (a beam with $\rho=1$) and beam constructed by tessellating UC1-UC4. All the lattice-based beams show a higher electrical to mechanical energy conversion compared to the solid beam. A higher electric potential is observed at the joints of struts close to the fixed end, due to higher bending stress and strain. UC1 and UC3 (Figs \ref{fig:Bending-direct} (c) and (e)) show higher energy conversion due to higher bending resulting from the transverse loading. Interestingly, UC4 (Fig \ref{fig:Bending-direct} (f)) shows a lower energy conversion, as compared to compression mode. This can be attributed to the fact that while in compression the horizontal strut in UC4 is under bending and experiences a higher strain gradient, while in beam-bending configuration, this effect is reduced.

To further investigate the size effects of flexoelectricity in truss-lattice materials, we compare the electromechanical coupling factor for solid beam and UC1-based truss-lattice beam for different thicknesses (Fig \ref{fig:Bending-converse} (g)). As the thickness of the solid beam increases, a rapid decrease in $K_{EM}$ is observed -- corroborating with the theory of flexoelectricity for beams. However, a very small decrease in $K_{EM}$ is observed in the UC1 lattice-based beam -- demonstrating a potential for not only achieving a high magnitude of $K_{EM}$ but also scaling up flexoelectricity through accumulated strain gradients in architected materials. 

\subsection{Converse flexoelectric effect}
\label{subsec:Converse_Flexo}
\begin{figure}[ht!]
    \centering
    \includegraphics[width=\textwidth]{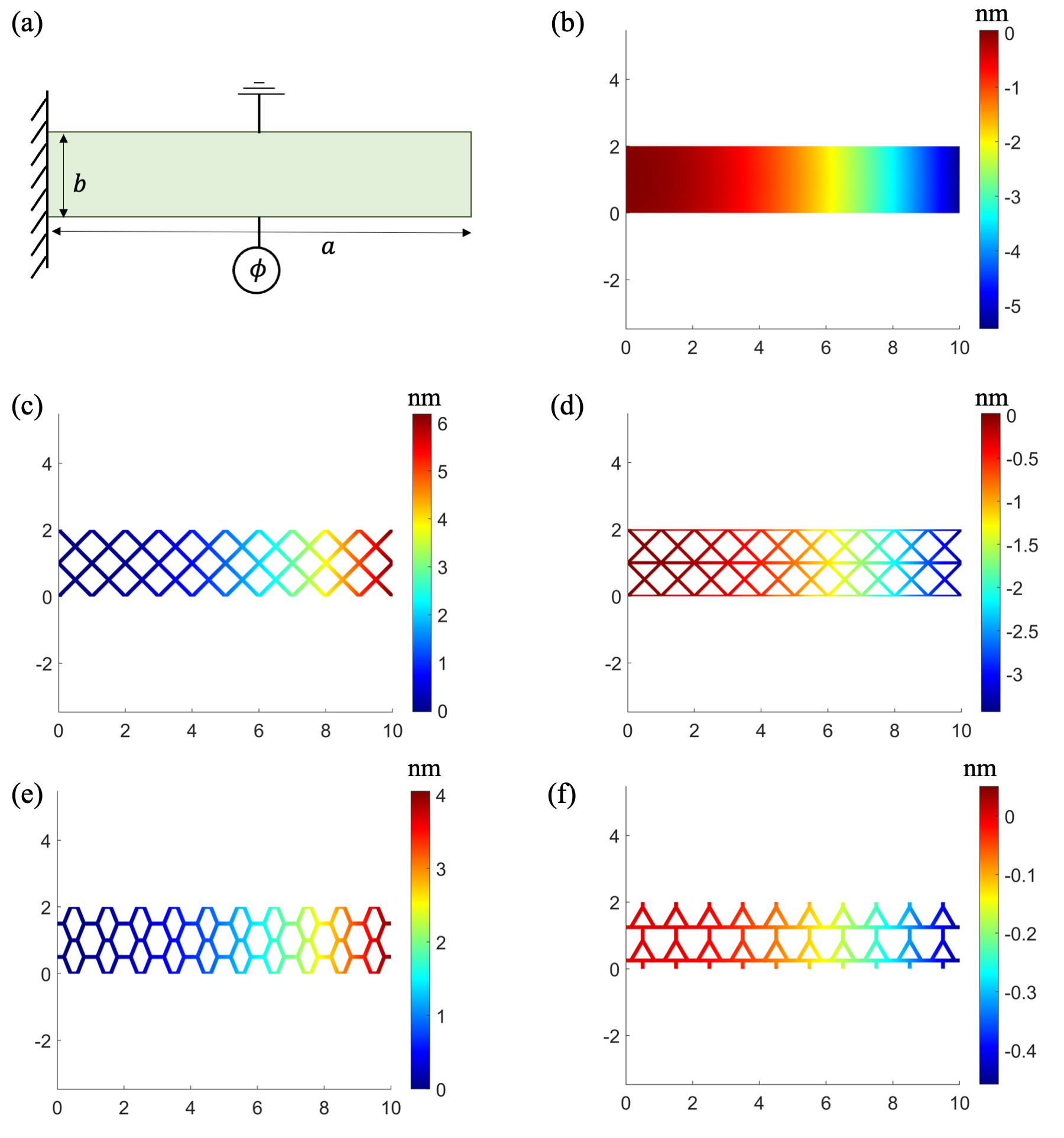}
    \caption{Converse flexoelectricity in a cantilever beam under an applied potential difference. \textbf{(a)} Geometry and boundary conditions of the beam, with $a=2 \mathrm{\mu m}$, $b=10 \mathrm{\mu m}$, and $\phi = 20$ V. \textbf{(b)} Displacement distribution in the solid beam and \textbf{(c)-(f)} UC1-UC4, under the applied electric potential difference. While the solid beam has a negative displacement, truss-lattices can have negative or positive displacement, depending on the electric field distribution varying with the type of microstructure.}
    \label{fig:Bending-converse}
\end{figure}

While the conversion of mechanical energy into electrical energy (sensing) in architected dielectric materials is crucial for different processes and applications such as bone regeneration and repair, hearing mechanism, and micro/nano sensors and energy harvesters; conversion of electrical energy into mechanical energy (actuation) for transmitting motion in architected materials can be pivotal in micro-robotics, deep brain stimulation, and many more applications. Here we investigate the electrical to mechanical energy conversion facilitated by the converse flexoelectric effect in truss lattice materials. A cantilever beam with width $a=20 \mu$m, and height $b=2 \mu$m (Fig \ref{fig:Bending-converse} (a)) is subjected to an electric field across its thickness and the vertical displacement ($u_y$) distribution is analyzed. Although the applied electric field would be uniform in the absence of flexoelectricity, due to the coupling of gradients of the mechanical and electrical fields, a non-uniform distribution of electric field across the thickness is generated. This gradient of electric field leads to a net electromechanical response in the form of bending. Fig \ref{fig:Bending-converse} (b) shows the displacement of a cantilever beam with $\rho=1$ (solid beam configuration) experiencing a maximum of 5.42 nm vertical displacement under an applied potential of $\phi = 20$ V. Figs. \ref{fig:Bending-converse} (c-f) display the displacement profile in the cantilever beam with truss lattice-based architectures based on UC1-UC4 respectively. While the solid beam shows a negative deflection, truss lattice beams experience negative or positive deflection, depending upon the microstructure of the beam, due to variations in the magnitude and direction of electric field gradients. Furthermore, while UC1 and UC3 beams show similar magnitudes of deflection as the solid beam, UC2 and UC4 experience relatively lower actuation. This can be attributed to the fact that while lattice beams experience higher magnitudes of electric field gradients, the fluctuations in sign of the same from positive to negative, lead to a reduction in the overall flexoelectric effect.

\section{Conclusion}
\label{sec:Conclusion}
An IGA-based multi-patch computational framework is presented for flexoelectricity in 2D truss-based architected materials. The requirement of $C^1$ continuity of the basis functions across patch boundaries is met in the weak form of the boundary value problem by implementing an interior penalty-based discontinuous Galerkin method. The method employs a stabilization or penalty term to ensure the stability of the solution as the value of the coefficient of this term (stabilization parameter $\tau$) is increased. The presented formulation is tested by evaluating the electromechanical coupling factors of flexoelectric and piezoelectric-flexoelectric beams discretized using multiple patches and validating against the analytical results. The present approach is found to be efficient in eliminating the jump in the second derivatives of the primary variables, observed with the $C^0$ continuous formulations. Further, through the simulation studies on 2D dielectric truss lattices -- based on four different \textit{unit cell} topologies, and their different tessellations --  the electromechanical response of micro-architected dielectric materials is investigated. Case studies for both direct and converse electromechanical couplings are conducted. In the direct effect -- i.e., generation of mechanical output in response to an electrical input -- the lattice-based structures are found to outperform the analogous solid structures. While the solid specimen would not yield any net electromechanical output under compression, due to uniform strain and zero strain gradient, the lattice materials are able to provide a net electromechanical output. Moreover, in bending mode, the lattice materials not only yield a higher electromechanical coupling factor but also show a lower rate of reduction in flexoelectricity with the increasing size of the specimen -- indicating a potential to scale up flexoelectric coupling from nano/micro scales to meso/macro scales. Further investigation is warranted to unveil the possibilities of designing metamaterials to obtain a significant level of flexoelectric response in meso/macro scale dielectrics.
\section*{Acknowledgement}
S.S. acknowledges the financial support provided by the \textit{Alexander von Humboldt foundation} through a postdoctoral fellowship.
\appendix
\section{Material and gradient matrices}
\label{sec:appendixA}
The elasticity, gradient elasticity, electrical, piezoelectric, and flexoelectric tensors are written in the matrix form for the 2D case as
{\allowdisplaybreaks
\begin{align}
    \bm{\dsC}& = \frac{E}{(1+\nu)(1-2\nu)}
        \begin{bmatrix}
        1-\nu & \nu & 0\\
        \nu & 1-\nu & 0 \\
        0 & 0 &  (1-2\nu)/2
    \end{bmatrix}, \notag \quad
    \bm{\mu} = \begin{bmatrix}
        \mu_{11} & \mu_{12} & 0 & 0 & 0 & \mu_{44}\\
        0 & 0 & \mu_{44} & \mu_{12} & \mu_{11} & 0
    \end{bmatrix}, \notag \\
    \bm{\kappa}& = 
    \begin{bmatrix}
        \kappa_{11} & 0 \\
        0 & \kappa_{22}
    \end{bmatrix}, \quad 
    \bfe=\begin{bmatrix}
        0 & 0 & e_{15} \\
        e_{21} & e_{22} & 0
    \end{bmatrix}, \ \text{and} \quad
    \bfh =
    \calL^2
    \begin{bmatrix}
    C_{11} & C_{12} & 0 & 0 & 0 & 0\\
    C_{12} & C_{11} & 0 & 0 & 0 & 0\\
    0 & 0 & C_{44} & 0 & 0 & 0 \\
    0 & 0 & 0 & C_{11} & C_{12} & 0\\
    0 & 0 & 0 & C_{12} & C_{11} & 0\\
    0 & 0 & 0 & 0 & 0 & C_{44}
    \end{bmatrix},
\end{align}
}
where $C_{11} = E(1-\nu)/(1+\nu)(1-2\nu)$, $C_{12} = E\nu/(1+\nu)(1-2\nu)$, $C_{44} = E(1-2\nu)/2(1+\nu)(1-2\nu)$, and $\calL$ is the length scale parameter taken as $0.1$ nm. The strain, strain gradient and electric field used in equation \eqref{eq:29}, are written in vector form as
\be
\bm{\epsilon}= 
\begin{bmatrix}
    \epsilon_{11} & \epsilon_{22} & \epsilon_{12}
\end{bmatrix}^T, \quad
\nabla\bm{\epsilon}=
\begin{bmatrix}
    \epsilon_{11,1} & \epsilon_{22,1} & \epsilon_{12,1} & \epsilon_{11,2} & \epsilon_{22,2} & \epsilon_{12,2}
\end{bmatrix}^T, \ \text{and} \quad 
\bfE=
\begin{bmatrix}
    E_1 & E_2
\end{bmatrix}^T.
\ee
The strain-displacement, electric field-potential, and the Hessian matrices are written as
{\allowdisplaybreaks
\begin{align}
\bfB_u&=
\begin{bmatrix}
    \frac{\partial R_1}{\partial x} & 0 & \frac{\partial R_1}{\partial y} \\
    0 & \frac{\partial R_1}{\partial y} & \frac{\partial R_1}{\partial x} \\
    \frac{\partial R_2}{\partial x} & 0 & \frac{\partial R_2}{\partial y} \\
    0 & \frac{\partial R_2}{\partial y} &  \frac{\partial R_2}{\partial x} \\
    \vdots & \vdots & \vdots \\
    \frac{\partial R_{n_{cp}}}{\partial x} & 0 & \frac{\partial R_{n_{cp}}}{\partial y} \\
    0 & \frac{\partial R_{n_{cp}}}{\partial y} & \frac{\partial R_{n_{cp}}}{\partial x}
\end{bmatrix}, \quad
\bfB_{\phi}=
\begin{bmatrix}
    \frac{\partial R_1}{\partial x} & \frac{\partial R_1}{\partial y} \\
    \frac{\partial R_2}{\partial x} & \frac{\partial R_2}{\partial y} \\
    \vdots & \vdots \\
    \frac{\partial R_{n_{cp}}}{\partial x} & \frac{\partial R_{n_{cp}}}{\partial y}
\end{bmatrix}, \notag \\
\bfH_u& = 
\begin{bmatrix}
    \frac{\partial^2 R_1}{\partial x^2} & 0 & \frac{\partial^2 R_1}{\partial y \partial x} & \frac{\partial^2 R_1}{\partial x \partial y} & 0 & \frac{\partial^2 R_1}{\partial y^2}\\
    0 & \frac{\partial^2 R_1}{\partial y \partial x} & \frac{\partial^2 R_1}{\partial x^2} & 0 & \frac{\partial^2 R_1}{\partial y^2} & \frac{\partial^2 R_1}{\partial x \partial y}\\
    \frac{\partial^2 R_2}{\partial x^2} & 0 & \frac{\partial^2 R_2}{\partial y \partial x} & \frac{\partial^2 R_2}{\partial x \partial y} & 0 & \frac{\partial^2 R_2}{\partial y^2}\\
    0 & \frac{\partial^2 R_2}{\partial y \partial x} & \frac{\partial^2 R_2}{\partial x^2} & 0 & \frac{\partial^2 R_2}{\partial y^2} & \frac{\partial^2 R_2}{\partial x \partial y}\\
    \vdots & \vdots & \vdots & \vdots & \vdots & \vdots \\
    \frac{\partial^2 R_{n_{cp}}}{\partial x^2} & 0 & \frac{\partial^2 R_{n_{cp}}}{\partial y \partial x} & \frac{\partial^2 R_{n_{cp}}}{\partial x \partial y} & 0 & \frac{\partial^2 R_{n_{cp}}}{\partial y^2}\\
    0 & \frac{\partial^2 R_{n_{cp}}}{\partial y \partial x} & \frac{\partial^2 R_{n_{cp}}}{\partial x^2} & 0 & \frac{\partial^2 R_{n_{cp}}}{\partial y^2} & \frac{\partial^2 R_{n_{cp}}}{\partial x \partial y} 
\end{bmatrix}, \ \text{and} \notag \\
\bfH_{\phi}& = 
\begin{bmatrix}
    \frac{\partial R_1}{\partial x} & 0 & \frac{\partial R_1}{\partial y} & 0\\
    0 & \frac{\partial R_1}{\partial x} & 0 & \frac{\partial R_1}{\partial y} \\
    \frac{\partial R_2}{\partial x} & 0 & \frac{\partial R_2}{\partial y} & 0\\
    0 & \frac{\partial R_2}{\partial x} & 0 & \frac{\partial R_2}{\partial y} \\
    \vdots & \vdots & \vdots & \vdots \\ 
    \frac{\partial R_{n_{cp}}}{\partial x} & 0 & \frac{\partial R_{n_{cp}}}{\partial y} & 0\\
    0 & \frac{\partial R_{n_{cp}}}{\partial x} & 0 & \frac{\partial R_{n_{cp}}}{\partial y}
\end{bmatrix}.
\end{align}
}
\setcounter{figure}{0}
\clearpage
\section{Convergence study and symmetric boundary conditions}
\label{sec:appendixB}
\begin{figure}[!htb]
    \centering
    \includegraphics[width=\textwidth, keepaspectratio]{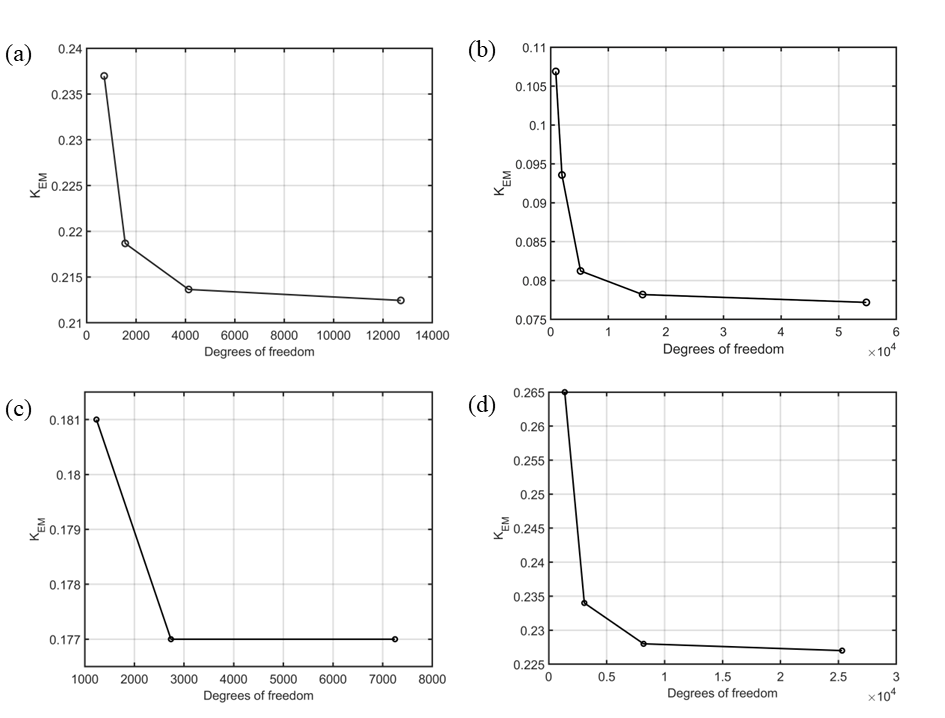}
    \captionsetup{justification=justified}
    \caption{Convergence of electromechanical coupling factor with mesh refinement for UC1-UC4.}
    \label{fig:UC_convergence}
\end{figure}

\begin{figure}[!htb]
    \centering
    \includegraphics[width=\textwidth, keepaspectratio]{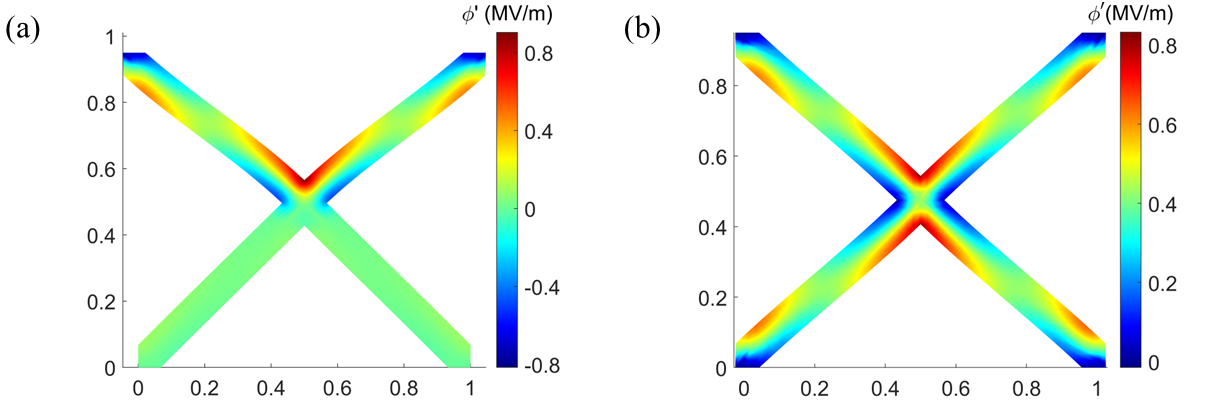}
    \captionsetup{justification=justified}
    \caption{Comparison of electric potential generated in UC1 under (a) compressive loads with bottom fixed, and (b) compressive load with roller supports at the bottom, facilitating symmetric strain distribution.}
    \label{fig:Symmetric_BC}
\end{figure}

 \bibliographystyle{elsarticle-num} 
 \bibliography{cas-refs}

\end{document}